\documentclass[aps,pra,twocolumn,superscriptaddress]{revtex4-1}
\usepackage{graphicx,ulem}
\usepackage[dvipsnames]{xcolor}
\usepackage{amsmath, amssymb}
\usepackage[colorlinks=true,urlcolor=blue,citecolor=blue,linkcolor=magenta]{hyperref}

\begin{document}

\title{Detection-device-independent verification of nonclassical light}

\author{Martin Bohmann}\email{martin.bohmann@ino.it}
	\affiliation{INO-CNR and LENS, Largo Enrico Fermi 2, I-50125 Firenze, Italy}
\author{Luo Qi}
	\affiliation{Max-Planck-Institute for the Science of Light, Erlangen, Germany}
	\affiliation{University of Erlangen-N\"urnberg, Staudtstrasse 7/B2, D-91058 Erlangen, Germany}
\author{Werner Vogel}
	\affiliation{Arbeitsgruppe Theoretische Quantenoptik, Institut f\"ur Physik, Universit\"at Rostock, D-18051 Rostock, Germany}
\author{Maria Chekhova}\email{maria.chekhova@mpl.mpg.de}
	\affiliation{Max-Planck-Institute for the Science of Light, Erlangen, Germany}
	\affiliation{University of Erlangen-N\"urnberg, Staudtstrasse 7/B2, D-91058 Erlangen, Germany}

\begin{abstract}
	The efficient certification of nonclassical effects of light forms the basis for applications in optical quantum technologies.
	We derive general correlation conditions for the verification of nonclassical light based on multiplexed detection.
	The obtained nonclassicality criteria are valid for imperfectly balanced multiplexing scenarios with on-off detectors and do not require any knowledge about the detector system.
	In this sense, they are fully independent of the detector system.
	In our experiment, we study light emitted by clusters of single-photon emitters, whose photon number may exceed the number of detection channels.
	Even under such conditions, our criteria certify nonclassicality with high statistical significance.

\end{abstract}

\date{\today}

\maketitle

\section{Introduction}	
	
	The verification of quantum correlations in optical systems is a key task in quantum optics.
	Besides its fundamental significance for the understanding of radiation fields, the identification of genuine quantum features is becoming ever more important as they can be used for applications in quantum technologies \cite{Braunstein05,Kok07,Gisin07}.
	A major goal, in this context, is to develop robust methods which ideally do not rely on any knowledge or assumptions about the studied system, leading  to the concept of device-independent quantum characterization \cite{Branciard13,Lim14,Gheorghiu15,Altorio16,Sperling17}.
	
	An important task is the characterization of light in the few-photon regime.
	For the analysis of quantum light in this regime, so-called multiplexing strategies \cite{Paul96,Kok01,Banaszek03,Rehavcek03,Fitch03,Achilles03,Castelletto07,Schettini07,Lundeen09,Matthews16} have been developed  as a way of gaining insights in the measured quantum state even when a photon-number-resolving measurement is not accessible. 
	Such strategies do not provide a direct access to the photon-number distribution and consequently the interpretation of the measurement statistics as the photon-number statistics can lead to a false certification of nonclassicality \cite{Sperling12}.
	However, nonclassicality criteria which can be directly applied to the recorded click-counting statistics have been formulated \cite{Sperling12a,Sperling13,Luis15,Filip13,Lachman16,Lee16}.
	In particular, such criteria are very efficient and successful in certifying nonclassicality from experimental data \cite{Bartley13,Sperling15,Heilmann16,Sperling16,Bohmann17b,Bohmann18,Obsil18,Tiedau18}.
	
	One common assumption for such conditions is that the incoming light is equally split and detected in each detection channel.
	Recently, the detector-independent verification of quantum light for such equal splitting has been reported \cite{Sperling17,Sperling17b}.
	In some cases, however, an equal splitting ratio might be hard to realize and requires the careful characterization of the optical elements.
	Furthermore, other multiplexing strategies such as fiber-loop detectors \cite{Banaszek03,Tiedau18} by design do not provide an equal splitting.
	For such unequal-splitting scenarios, a condition based on second-order moments \cite{Lee16} has been derived as a generalization of the corresponding equal-splitting condition \cite{Sperling12a}.
	More general higher order criteria have, however, not yet been reported for such unequal-splitting detections.
	
	In this paper, we introduce detector-independent general (higher order) conditions for the certification of nonclassical light measured with unbalanced multiplexing schemes and on-off detectors.
	The presented conditions are fully independent of the properties of the used detection scheme.
	Based on Chebyshev's integral inequality, we derive a family of inequality conditions for the no-click events at the output channels which have to be fulfilled for any classical radiation fields; their violation verifies nonclassicality.
	The so-obtained inequalities include simple covariance conditions between two detection channels and more general higher order correlation conditions.
	Our approach is based on minimal assumptions and requirements which guarantees the applicability to any multiplexing setup, even without the knowledge about the used detectors and the splitting ratios.
	We demonstrate the strength of the obtained criteria by certifying nonclassicality of light from clusters of single-photon emitters with high statistical significance.
	The relations of the presented nonclassicality certifiers to other nonclassicality criteria based on the Mandel $Q$ parameter and the matrix of moments approach are discussed.

\section{Multiplexing detection}

\begin{figure}[h]
		\centering
		\includegraphics[width=0.75\columnwidth]{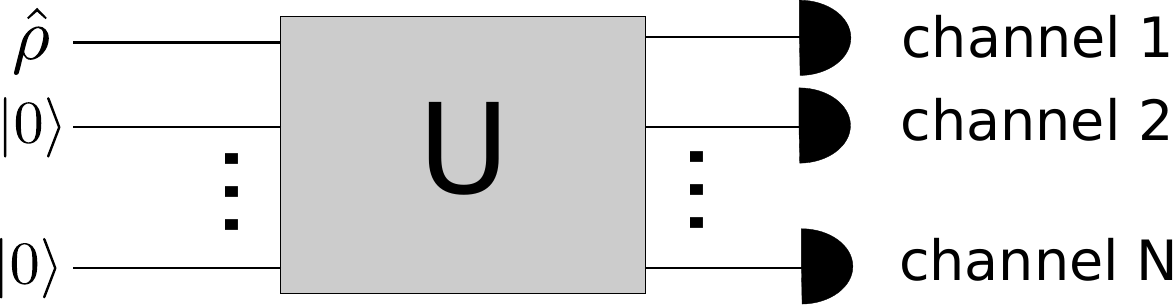}
			\caption{
				Working principle of a multiplexing device.
				The input quantum state $\hat \rho$ is split at an unbalanced multiport splitter and each output channel is measured with an on-off detector.
			}
		\label{fig:Multiplexing}
	\end{figure}

	We are interested in the certification of quantum correlations based on general multiplexing scenarios, as schematically sketched in Fig. \ref{fig:Multiplexing}.
	The incident quantum state of light is (unequally) split into $N$ output channels.
	Each of these channels is then measured by a single on-off detector, which is the standard working principle of multiplexing detectors \cite{Paul96,Kok01,Banaszek03,Rehavcek03,Fitch03,Achilles03,Castelletto07,Schettini07,Lundeen09,Matthews16}.
	
	Let us describe this multiplexing step formally.
	We express the input quantum state in terms of the Glauber Sudarshan $P$ representation \cite{Glauber63,Sudarshan63},
	\begin{align}
		\hat \rho_{\mathrm{in}}=\int d^2\alpha P(\alpha)|\alpha\rangle\langle\alpha|,
	\end{align}
	where $|\alpha\rangle$ is a coherent state.
	A quantum state is called classical if and only if (iff) its $P$ function is non-negative \cite{Titulaer65,Mandel86}.
	Multiplexing devices act as $N{\times} N$ multimode splitter and can be described via the unitary operation $\mathbf{U}(N)= (u_{i,j})_{i,j=1}^{N}$ which relates the input to the output field operators via $\hat{\mathbf{a}}_{\mathrm{out}}{=}\mathbf{U}(N)\hat{\mathbf{a}}_{\mathrm{in}}$ and $\hat{\mathbf{a}}_{\mathrm{in(out)}}{=}(\hat a_{1\mathrm{in(out)}},\dots,\hat a_{N\mathrm{in(out)}})$.
	In the multiplexing case, only the first mode is occupied and the other ones are in the vacuum state.
	Consequently, the output quantum state can be written as
	\begin{align}\label{eq:Pout}
		\hat \rho_{\mathrm{out}}{=}\int d^2\alpha P(\alpha)|u_{1,1}\alpha,\dots,u_{1,N}\alpha\rangle\langle u_{1,1}\alpha,\dots,u_{1,N}\alpha|,
	\end{align}
	with $\sum_k |u_{1,k}|^2{=}1$ \cite{Note1}.
	We explicitly do not restrict our consideration to the case of uniform splitting, i.e., $\forall k:\quad|u_{1,k}|^2=1/N$.
	
	We are now interested in the detector-click probability in each channel.
	The probability of detecting no click in the $k$-th channel is given by the expectation value 
	\begin{align}
	\langle{:} \hat m_k {:}\rangle=\int d^2\alpha P(\alpha)\langle u_{1,k}\alpha|{:}\hat m_k{:}|u_{1,k}\alpha\rangle,
	\end{align}
	where ${:}\dots {:}$ denotes the normal-ordering prescription; cf. \cite{Vogel06}.
	The corresponding operator is defined as $\hat m_{k}{=}e^{-\hat \Gamma_k(\hat n_k)}$,
	the subscript $k$ indicates the detection channel and $\hat n_k$ is the photon-number operator in the corresponding channel.
	The probability of obtaining a click in this channel is given by $1{-}\langle{:}\hat m_k{:}\rangle$.
	The detector response function $\hat \Gamma_k(\hat n_k)$ is a function of $\hat n_k$ and describes the connection between the electromagnetic field and the generation of a click \cite{Kelly64,Vogel06}.
	The detector response function can be determined via direct calibration techniques; see, e.g., \cite{Bohmann17a}.

\section{Conditions for quantum correlations}

	We aim at formulating nonclassicality conditions based on the correlations between the no-click events of the different output channels which do not depend on the characteristics of the used multiplexing architecture and detectors.
	The simplest case we can consider in this context is the correlation between the no-click events of two output channels \cite{Note2}.
	In fact, we can use  Chebyshev's integral inequality (see, e.g., \cite{Mitrinovic70}) to derive the simple condition,
	\begin{align}\label{eq:covar}
		\langle{:}\mathrm{Cov}(\hat m_i, \hat m_j){:}\rangle=\langle {:} \hat m_i \hat m_j {:} \rangle - \langle {:} \hat m_i {:}\rangle \langle {:} \hat m_j {:}\rangle \stackrel{\mathrm{cl}}\geq 0,
	\end{align}
	which must hold for any classical input state, i.e., a quantum state with a non-negative $P$ function.
	Details on the derivation are provided in the Appendix \ref{sec:derivation}.
	The violation of this inequality is a direct and experimentally easily accessible signature of nonclassicality and is directly related to the negativities of the $P$ function of the studied state.
	In particular, only one multiplexing step and on-off detection is sufficient for the application of condition \eqref{eq:covar}.
	
	Condition \eqref{eq:covar} has a clear physical interpretation.
	Non-negative normal-ordered covariances can be explained in terms of a classical description of the measured radiation field, i.e., by a classical $P$ function.
	In particular, for an input coherent state, the no-click events are uncorrelated [$\langle{:}\mathrm{Cov}(\hat m_i, \hat m_j){:}\rangle{=}0$], which represents the boundary between classical and nonclassical radiation fields.
	On the other hand, an anticorrelation, i.e., a negative covariance of the no-click events, can only arise from negativities in the $P$ function of the considered state.
	For example, a single-photon input state leads to an anticorrelation of the no-click events which is revealed by a negative covariance.
	
	We can further generalize condition \eqref{eq:covar} to multimode higher order moment conditions.
	Again by making use of Chebyshev's integral inequality \cite{Mitrinovic70}, we formulate the family of correlation conditions (see Appendix \ref{sec:derivation} for details)
	\begin{align}\label{eq:partitions}
		\langle {:} \hat m_{{\mathcal{I}}_1} \dots \hat m_{{\mathcal{I}}_K} {:}\rangle - \langle {:} \hat m_{{\mathcal{I}}_1} {:}\rangle \cdots \langle {:} \hat m_{{\mathcal{I}}_K} {:}\rangle \stackrel{\mathrm{cl}}\geq 0,
	\end{align}
	where $\mathcal{I}_1\dots\mathcal{I}_{K}$ are mutually disjoint subsets (partitions) of $\mathcal{I}=\{1,\dots,N\}$ and $\hat m_J$ is the no-click operator for all detection channels in $\mathcal{I}_J$, $\hat m_{\mathcal{I}_J}{=}\prod_{j\in \mathcal{I}_J}\hat m_j$.
	These general conditions also include asymmetric partitions and the clustering of channels.
	Note that Eq. \eqref{eq:partitions} generalizes the approach in \cite{Lachman16} to unequal splitting, uncharacterized detectors, and arbitrary partitions.
	In Appendix \ref{sec:derivation}, we show that the conditions \eqref{eq:covar} and \eqref{eq:partitions} are not affected by dark counts or other uncorrelated noise.

\section{Example}
	Before turning to the experiment and the data analysis, let us consider an example.
	As an input state, we choose an $n$-photon-added thermal state $\mathcal{N}_n\left(\hat a^\dagger\right)^n\hat\rho_{\mathrm{th}}\hat a^n$,  where the thermal state is $\hat\rho_{\mathrm{th}}=1/(\overline{n}+1)\sum_{k=0}^\infty \left(\overline{n}/(\overline{n}+1)\right)^k |k\rangle\langle k|$ with $\mathcal{N}_n$ being the normalization constant \cite{Agarwal92}.
	Such states have been realized in experiments \cite{Zavatta04,Kiesel08}.
	We consider a single unbalanced multiplexing step (beam splitter) with an intensity splitting of $70{:}30$ and a detection efficiency of $0.7$.
	By violating the covariance inequality \eqref{eq:covar}, we can certify nonclassicality.
	We compare this nonideal detection scheme with a classicality condition based on the photon-number covariance, 
	\begin{align}\label{eq:ncovar}
		\langle \hat n_i\hat n_j \rangle-\langle \hat n_i \rangle\langle \hat n_j \rangle\stackrel{\mathrm{cl}}\geq 0,
	\end{align}
	where $\hat n_j$ is the photon-number operator in the $j$th detection channel.
	The application of this condition would require experimental access to the second-order moments of the photon-number operator.
	The violation of this condition corresponds to the nonclassicality condition in terms of the second-order intensity correlation function, $g^{(2)}(0)<1$, which is closely related to the sub-Poissonian photon statistics and the Mandel $Q$ parameter \cite{Mandel79}.
	
	\begin{figure}[t]
		\centering
		\includegraphics[width=0.8\columnwidth]{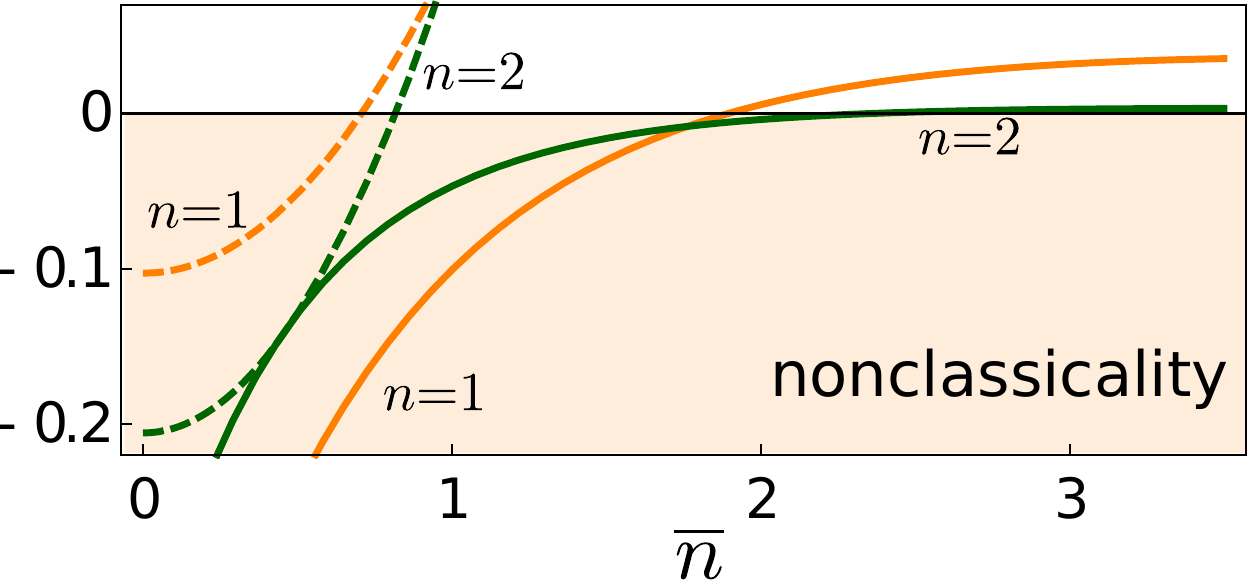}
			\caption{
				The no-click-covariance condition \eqref{eq:covar} (solid lines; scaled by a factor of $5$) and the photon-number covariance [Eq. \eqref{eq:ncovar}] (dashed) are shown for one- and two-photon-added thermal states in dependence on the mean thermal photon number $\overline{n}$.
			}
		\label{fig:mpats}
	\end{figure}
	
	In Fig. \ref{fig:mpats}, the behavior of the two conditions is shown for one and two-photon-added thermal states in dependence on the thermal photon number.
	The violation of the no-click inequality \eqref{eq:covar} detects nonclassicality in a wider  range of $\overline{n}$ than corresponding violation of the photon-number condition \eqref{eq:ncovar}.
	Similar behaviors have been observed for the sampling of phase-space distributions from multiplexing detection data \cite{Luis15,Bohmann18}.
	
	An explanation for this behavior can be found when one considers the measurement operator of the click detection, i.e., the no-click operator $\hat m_k$.
	This operator is an exponential function of the photon-number operator and, thus, higher order moments of the photon-number operator contribute to the condition~\eqref{eq:covar}.
	Therefore, the click detection may be more sensitive toward nonclassical effects than the detection of the first two moments of the photon-number operator in Eq. \eqref{eq:ncovar}.

\section{Experiment}
	In our experimental setup, we study light from clusters of single-photon emitters which we detect with the help of time-bin multiplexed click detection.
	We used multiphoton light emitted by clusters of colloidal CdSe/CdS quantum ''dot in rods'' (DRs)~\cite{Carbone2007,Pisanello2010,Manceau2014,Vezzoli2015} coated on a fused silica substrate and excited by picosecond pulses at $355$ nm. 
	For a detailed description of the experiment, see Ref.~\cite{Qi2018}.
	To get rid of the pump radiation, the emitted light was filtered using a long-pass filter and a band-pass filter (center wavelength $607$ nm, bandwidth $42$ nm). 
	The size of the cluster was determined by assuming that the mean number of photons emitted per pulse scales with the number of emitters in the cluster. 
	In this way, we obtained clusters with an effective size between $2$ and $14$ emitters. 
	Each DR in such a cluster, provided that it is excited, emits a quantum state that is close to a single-photon one, with an extremely small admixture of a two-photon component. 
	Taking into account the $25\%$ excitation probability per excitation pulse, the state emitted by a single DR can be written as 
	$\hat \rho_{\mathrm{DR}}{=}p_0|0\rangle\langle0|+p_1|1\rangle\langle1|+p_2|2\rangle\langle2|$,
	where $p_0{\approx}0.9$, $p_1{\approx}0.1$, and $p_2{\approx}10^{-4}$. 
	The low probability of two-photon emission leads to strongly nonclassical $g^{(2)}(0){\le}0.05$.
	Although different DRs in a cluster emit incoherently, the resulting state  manifests nonclassicality because the total number of emitted photons is restricted according to the size of the cluster~\cite{Shcherbina2014,Qi2018}.
	
\begin{figure}[t]
		\centering
		\includegraphics[width=0.9\columnwidth]{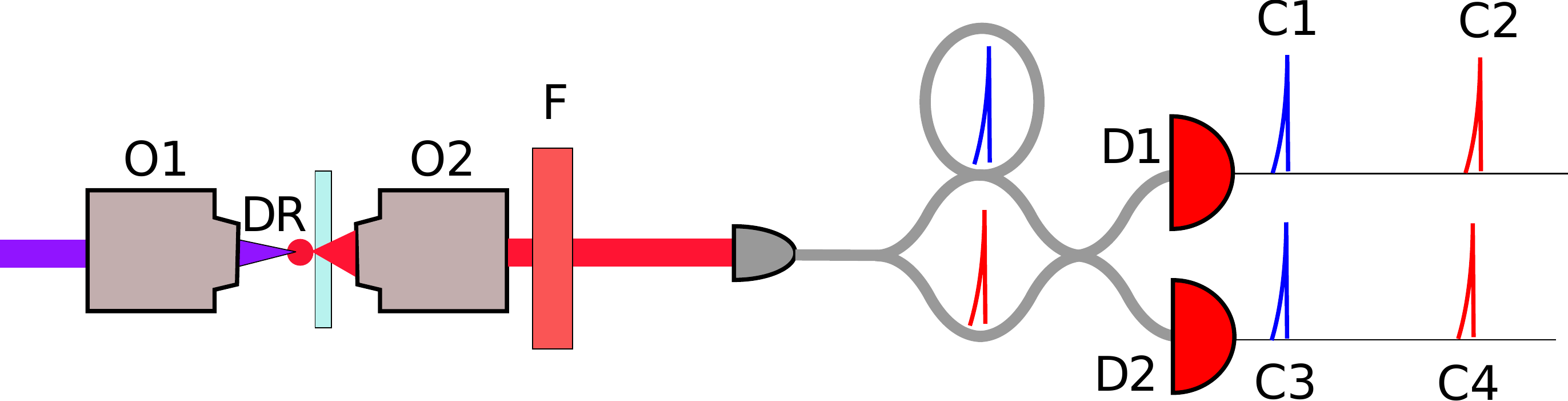}
			\caption{
				Principal scheme of the setup. The pump radiation ($355$ nm) is focused into a cluster of DRs through objective lens O1 and then cut off by the filters F. 
				The radiation emitted by the cluster is coupled into an objective lens O2 and, after filtering, is sent into the fiber-assisted multiplexed detection setup where two detectors D1, D2 and two different path lengths create four detection channels, C1, $\dots$, C4.
			}
		\label{fig:setup}
	\end{figure}

	The radiation emitted by a single cluster was collected with an efficiency of $44\%$, which takes into account the losses at all optical elements, and sent to a fiber multiplexed detection setup, cf. Fig.~\ref{fig:setup}, comprising two click detectors based on avalanche photodiodes, with the quantum efficiency $60\%$. 
	The use of fiber loops provided two time bins for each detector, and therefore the setup was equivalent to the one shown in Fig.~\ref{fig:Multiplexing}, with $N{=}4$ channels. 
	For each of the studied clusters, we collected a dataset containing between $10^7$ and $10^8$ pulses, and for each pulse, the number of click counts in each channel was registered. 
	Depending on the size of the cluster, the mean click number per pulse was between 0.05 and 0.1 due to the low excitation and detection efficiency. 

\section{Results}
	We apply the nonclassicality criteria based on the inequalities \eqref{eq:covar} and \eqref{eq:partitions} to the multiplexing data obtained for the different cluster sizes.
	Our approach can be directly applied to the measured data without the knowledge of the detection system or data post-processing.
	We analyze the correlation condition \eqref{eq:covar} between two of the detection channels (first and third) and the condition corresponding to the full partition of all four detection channels [cf. Eq. \eqref{eq:partitions}] for the different cluster sizes.
	The no-click moments and their statistical errors can be directly sampled from the measurement statistics \cite{Sperling15}.
	
	\begin{figure}[t]
		\centering
		\includegraphics[width=0.8\columnwidth]{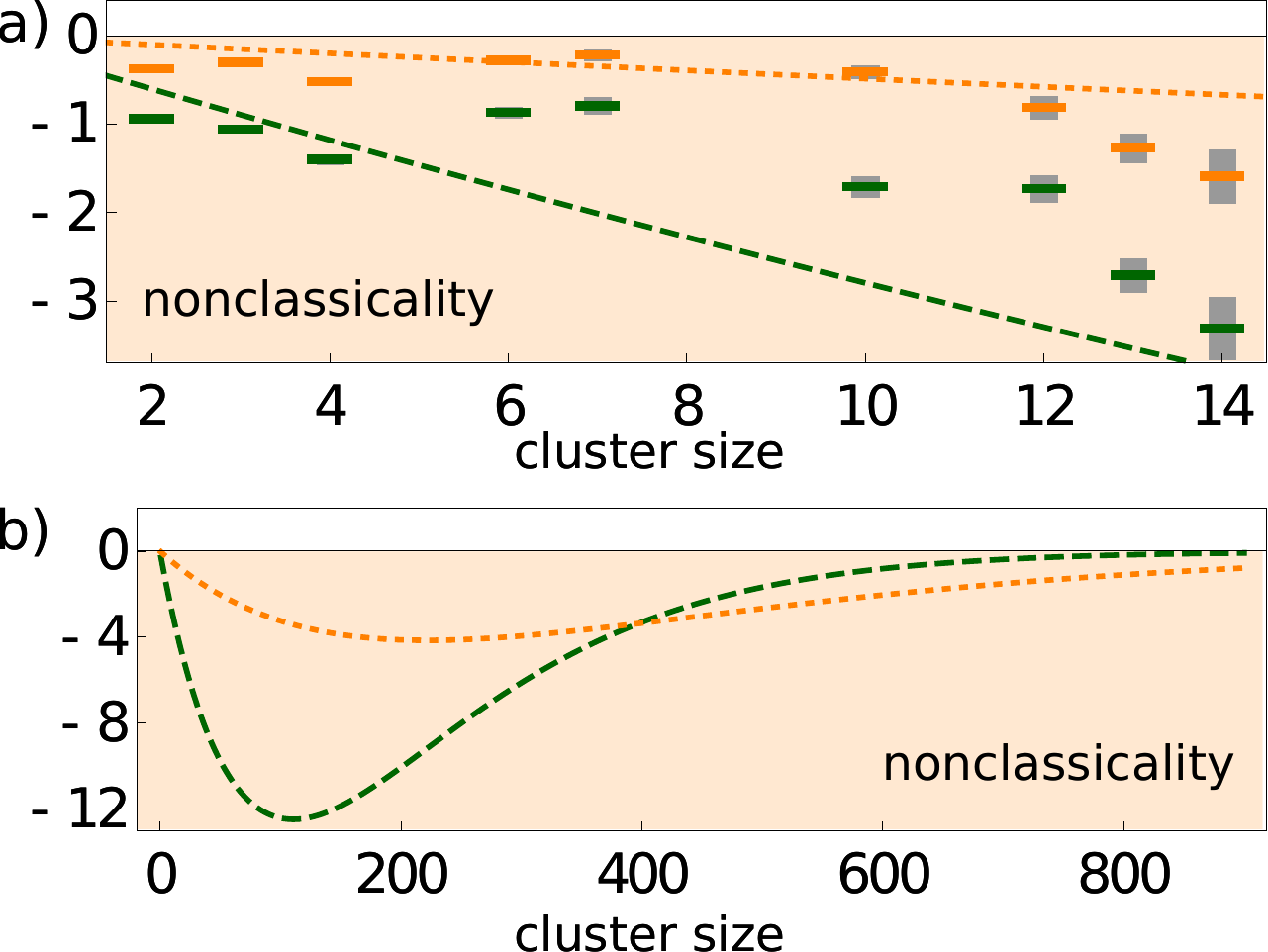}
			\caption{
				a) The covariance condition \eqref{eq:covar} between two detection channels (upper, orange markers) and the condition based on the full partition of all channels, cf. Eq. \eqref{eq:partitions} (lower, green markers), scaled by a factor of $10^4$, are shown together with their sampling errors (gray bars) for different cluster sizes.
				The dashed and dotted lines represent the corresponding model, Eqs. \eqref{eq:model1} and \eqref{eq:model2} with $\eta=0.009$ and different $M$ (the lines should guide the eye).
				b) The behavior for higher numbers of emitters following the theoretical model.
				The nonclassicality conditions only approach the classical limit for high number of emitters due to saturation of all detectors.
			}
		\label{fig:data1}
	\end{figure}
	
	The results are shown in Fig. \ref{fig:data1} a).
	Even the simple covariance condition \eqref{eq:covar} between two of the output channels is capable of certifying nonclassicality with statistical significances (estimated value divided by its statistical sampling error) above three standard derivations for all cluster sizes.
	Furthermore, we identify that the condition based on the full partition of the no-click operator into four parts [cf. Eq. \eqref{eq:partitions}] even yields verifications with higher statistical significances of up to $29.3$ standard deviations.
	Thus, the used conditions provide robust and easily applicable approaches for the faithful detector-independent verification of nonclassical light.
	Importantly, we can certify nonclassicality from the measurement outcomes of an uncharacterized detection device even if no assumption on the quantum state of the measured light is used.
	
	Let us discuss the dependence on the cluster size.
	We observe that for both conditions the distance to the classical limit (zero) increases with the number of emitters.
	At first glance, this is a surprising observation as the nonclassicality of the recorded light stems from the single-photon characteristics (''antibunching'') of the light from individual emitters.
	Therefore, one would expect that increasing the number of incoherently emitting quantum dots would suppress the nonclassical character and that the nonclassicality conditions eventually approach the classical limit.
	However, the relatively low efficiency (excitation, collection, detection) of the whole experiment leads to an initial increase of the distance to the classical limit with increasing numbers of emitters.
	
	The following model can explain this.
	We assume that each excited quantum dot emits a single photon and all losses and the non-unit excitation probability can be summarized by an overall quantum efficiency $\eta$.
	The light emitted by a cluster of $M$ incoherently emitting quantum dots is an $M$-mode tensor-product state of single-photon states.
	Furthermore, we assume for simplicity symmetric multiplexing, and that the $M$-mode state is mode-insensitively recorded by the detectors.
	This yield the expressions of the conditions \eqref{eq:covar} and \eqref{eq:partitions},
	\begin{align}
		&\langle{:}\mathrm{Cov}(\hat m_1, \hat m_3){:}\rangle=\left(1-\frac{\eta}{2}\right)^M-\left(1-\frac{\eta}{4}\right)^{2M}\quad \text{and}\label{eq:model1}
		\\
		&\langle {:} \hat m_{1} \dots \hat m_{4} {:}\rangle - \prod_{i=1}^{4} \langle {:} \hat m_{i} {:}\rangle= \left(1-\eta\right)^M-\left(1-\frac{\eta}{4}\right)^{4M}\label{eq:model2}
	\end{align}
	respectively; cf. Appendix \ref{sec:model} for more details.
	This simple model captures the dependence on the cluster size [Fig. \ref{fig:data1} a)] and explains that for the relatively low efficiency of the experiment ($\eta\approx0.009$) both conditions become more negative with increasing cluster size in the considered interval.
	Additionally, the model can be used to predict the further behavior for higher numbers of emitters [see Fig. \ref{fig:data1} b)], where we in fact observe that for a large number of emitters the conditions approach the classical limit.
	The overall efficiency $\eta$ determines how fast the classical limit is reached.
	Examples with other efficiencies are provided in Appendix \ref{sec:model}.
	Furthermore, we can conclude that with the proposed criteria it would be possible to certify nonclassicality of the light emitted by clusters of several hundreds of emitters with the used experimental setup.

\section{Discussion}

	The obtained covariance condition \eqref{eq:covar} is closely connected to the Mandel $Q$ parameter \cite{Mandel79} and related second-order moment conditions such as the sub-binomial $Q_{\mathrm{B}}$ \cite{Sperling12a} and sub-Poisson-binomial $Q_{\mathrm{PB}}$  \cite{Lee16} parameters.
	In fact, we show in Appendix \ref{sec:relation} that these criteria can be traced back to condition \eqref{eq:covar}.
	Thus, we could identify the covariance condition \eqref{eq:covar} as the fundamental building block of these other nonclassicality criteria.
	
	Let us now consider conditions including higher order moments, such as in Eq. \eqref{eq:partitions}.
	This includes also asymmetric partitions of the corresponding no-click operators, i.e., partitions in which different parts may have different number of elements, such as, e.g., the partition $\langle {:} \hat m_{1}\hat m_{2} {:}\rangle \langle {:} \hat m_{3} {:}\rangle$.
	Other approaches \cite{Sperling13,Sperling15,Heilmann16,Sperling17} are based on the matrix of moments which, by construction, cannot involve asymmetric-partition conditions.
	In Appendix \ref{sec:relation}, a comparison of these methods with the derived conditions is presented.
	Hence, the here derived conditions are by construction different from already existing approaches and provide a wider applicability \cite{Note3}.
	
	A detector-independent method for the certification of quantum light through multiplexing was already introduced \cite{Sperling17}.
	In that work, the detector independence refers solely to the detectors and not to the whole detection system, including the multiplexing, as an equal splitting into the detection channels is required.
	With our approach, we can relax the latter requirement and, hence, have a condition which is fully independent of the whole detection scheme.
	
	Furthermore, we would like to discuss the relation of the introduced nonclassicality conditions to entanglement.
	The derived conditions can certify nonclassicality of the input quantum state of a multiplexing device.
	In Sec. IV of Ref. \cite{Vogel14} it was undoubtedly demonstrated that the multiplexing of a nonclassical state of light yields multipartite entangled states. 
	This is just what happens in our experiment. However, the direct certification of the entanglement requires an extension of the measurement setup, which is beyond the scope of this paper.
	Although we are testing nonclassicality of the input state, our conditions in fact reveal quantum correlations between the different detection channels.
	Hence, the obtained nonclassicality conditions are closely related to multimode entanglement.
	This opens possibilities for future applications in quantum technologies.

\section{Conclusions}
	
	We formulated conditions for the verification of nonclassical light detected by general multiplexing setups.
	The obtained criteria do not require any knowledge about splitting ratios and used detectors.
	Hence, our methods are fully independent of the detection scheme.
	The obtained correlation criteria include conditions based on second-order and higher order moments of the no-click events.
	We demonstrated the strength of our approach by certifying nonclassicality of light emitted by clusters of single-photon emitters.
	Importantly, the presented criteria are capable of detecting quantum light even if the number of photon emitters is higher than the number of used detection channels.
	
	We could show that our conditions based on on-off detectors can provide even more insight into the nonclassical character of the recorded light than comparable approaches based on ideal photon-number-resolving measurements.
	Furthermore, we discussed the relation to established nonclassicality indicators and showed that our approach provides new forms of nonclassicality conditions which cannot be deduced from other existing criteria.
	The present results provide useful and simple tools for the detector-independent verification of quantum light, applicable to many experimental scenarios.

\begin{acknowledgments}
	This work has been supported by Deutsche Forschungsgemeinschaft through Grant No. VO 501/22-2.
	MB acknowledges financial support by the Leopoldina Fellowship Programme of the German National Academy of Science (LPDS 2019-01) and thanks Valentin Gebhart and Elizabeth Agudelo for helpful comments.
\end{acknowledgments}

\appendix
\begin{widetext}
	\section{Derivation of the nonclassicality conditions}\label{sec:derivation}
	Here we will show how the introduced nonclassicality conditions can be derived by using Chebyshev's integral inequality.
	Furthermore, we demonstrate that the results of the obtained conditions are not influenced by uncorrelated noise sources.
	
\subsection{Chebyshev's integral inequality}
	In order to construct the nonclassicality conditions we will make use of Chebyshev's integral inequality; see, e.g.,~\cite{Mitrinovic70}.
	Let $f$ and $g$ be two functions which are integrable and monotone in the same sense on ($a,b$) and let $p$ be a positive and integrable function on the same interval. 
	Then the Chebyshev's integral inequality
	\begin{align}\label{eq:Chebyshev}
		\int_a^b p(x) f(x) g(x)dx \int_a^b p(x) dx \geq
		\int_a^b p(x) f(x)dx  \int_a^b p(x) g(x)dx,
	\end{align}
	holds.
	In the following, we will see that we can make use of this inequality for the derivation of the nonclassicality conditions.
	In our case, the $p(x)$ will be the (phase-averaged) $P$ function of a classical quantum state and $f$, $g$ will be the expectation values of the normal-ordered no-click operators with coherent states.

\subsection{Two-channel no-click correlation}
	The starting point of this consideration is the multimode state after the multiplexing step,
	\begin{align}
		\hat \rho_{\mathrm{out}}=\int d^2\alpha P(\alpha)|u_{1,1}\alpha,\dots,u_{1,N}\alpha\rangle\langle u_{1,1}\alpha,\dots,u_{1,N}\alpha|,
	\end{align}
	with $\sum_k |u_{1,k}|^2=1$.
	We consider the no-click operators $\hat m_i=e^{-\Gamma(\hat n_i)}$ whose normally ordered expectation values with coherent states, $\langle \beta| {:}\hat m_i|\beta{:}\rangle=\langle \beta| {:}e^{-\Gamma(\hat n_i)} {:}|\beta\rangle=e^{-\Gamma(|\beta|^2)}$, are monotonically decreasing functions of $|\beta|$.
	A typical example of a detector response function is a linear response function $ \Gamma=\eta_i|\beta|^2+\nu_i$, where $\eta_i$ and $\nu_i$ are the quantum efficiency and the dark-count rate in the $i$th mode, respectively.
	
	Then, the normal-ordered expectation value of the product of the two no-click operators $\hat m_i$ and $\hat m_j$ reads as
	\begin{align}
		\langle{:}\hat m_i \hat m_j{:}\rangle=\mathrm{Tr}[\hat \rho_{\mathrm{out}}{:}\hat m_i \hat m_j{:}]=\int d^2\alpha P(\alpha) e^{-\Gamma_i(|u_{1,i}|^2|\alpha|^2)}e^{-\Gamma_j(|u_{1,j}|^2|\alpha|^2)}.
	\end{align}
	As both functions $e^{-\Gamma(|u_{1,i}|^2|\alpha|^2)}$ and $e^{-\Gamma(|u_{1,i}|^2|\alpha|^2)}$ are monotonically decreasing functions of $|\alpha|^2$ and we assume a non-negative (classical) $P$ distribution, we can apply Chebyshev's integral inequality and obtain
	\begin{align}
		\int d^2\alpha P(\alpha) e^{-\Gamma_i(|u_i|^2|\alpha|^2)}e^{-\Gamma_j(|u_j|^2|\alpha|^2)} \geq \int d^2\alpha P(\alpha) e^{-\Gamma_i(|u_i|^2|\alpha|^2)} \int d^2\alpha P(\alpha) e^{-\Gamma_j(|u_j|^2|\alpha|^2)},
	\end{align}
	which can be written in terms of the covariance,
	\begin{align}
		\langle{:}\mathrm{Cov}(\hat m_i, \hat m_j){:}\rangle=\langle{:}\hat m_i \hat m_j{:}\rangle - \langle{:}\hat m_i {:}\rangle\langle{:} \hat m_j{:}\rangle \stackrel{\mathrm{cl}}\geq 0.
	\end{align}
	This inequality has to hold for any non-negative (classical) $P$ and, thus, a violation of the inequality immediately uncovers nonclassicality of the considered state.
	Let us stress that the derivation does not rely on the knowledge of the splitting ratios and the properties of the detectors.

\subsection{Multimode generalization}
	Here, we will show how the multimode generalizations of the two-mode covariance condition can be obtained by applying Chebyshev's integral inequality several times. 
	To derive the multimode conditions for $N$ detection modes, we make several times use of Chebyshev's integral inequality.
	We start from the expectation value 
	\begin{align}
	\langle{:}\prod_{i=1}^N\hat m_i {:}\rangle=\int d^2\alpha P(\alpha) \prod_{i=1}^N e^{-\Gamma_i(|u_{1,i}|^2|\alpha|^2)},
	\end{align}
	which can be written as
	\begin{align}
	\langle{:}\prod_{i=1}^N\hat m_i {:}\rangle=\int d^2\alpha P(\alpha) e^{-\sum_{i=1}^N\Gamma_i(|u_{1,i}|^2|\alpha|^2)}=\int d^2\alpha P(\alpha) e^{-\sum_{i\in \mathcal{I}}\Gamma_i(|u_{1,i}|^2|\alpha|^2)}e^{-\sum_{i\in \mathcal{\overline{I}}}\Gamma_i(|u_{1,i}|^2|\alpha|^2)}
	\end{align}
	where $\mathcal{I}$ and $ \mathcal{\overline{I}}$ are two bipartitions of the considered operator functions.
	Note that both $e^{-\sum_{i\in \mathcal{\overline{I}}}\Gamma_i(|u_{1,i}|^2|\alpha|^2)}$ and $e^{-\sum_{i\in \mathcal{I}}\Gamma_i(|u_{1,i}|^2|\alpha|^2)}$ are monotonically decreasing functions.
	As in the case above, we can now apply Chebyshev's integral inequality which yields
	\begin{align}
	\int d^2\alpha P(\alpha) e^{-\sum_{i=1}^N\Gamma_i(|u_{1,i}|^2|\alpha|^2)}\geq \int d^2\alpha P(\alpha) e^{-\sum_{i\in \mathcal{I}}\Gamma_i(|u_{1,i}|^2|\alpha|^2)} \int d^2\alpha P(\alpha) e^{-\sum_{i\in \mathcal{\overline{I}}}\Gamma_i(|u_{1,i}|^2|\alpha|^2)},
	\end{align}
	which holds for any non-negative $P$ function.
	This may also be written as $\langle{:}\prod_{i=1}^N\hat m_i {:}\rangle\geq \langle{:}\prod_{i\in \mathcal{J}}\hat m_i{:}\rangle\langle {:}\prod_{j\in \mathcal{\overline{J}}}\hat m_j{:}\rangle$.
	This procedure can be repeated several times which leads to the general form
	\begin{align}
		\langle {:} \hat m_{{\mathcal{I}}_1} \dots \hat m_{{\mathcal{I}}_K} {:}\rangle - \langle {:} \hat m_{{\mathcal{I}}_1} {:}\rangle \cdots \langle {:} \hat m_{{\mathcal{I}}_K} {:}\rangle \stackrel{\mathrm{cl}}\geq 0,
	\end{align}
	where $\mathcal{I}_1\dots\mathcal{I}_{K}$ are mutually disjoint subsets (partitions) of $\mathcal{I}=\{1,\dots,N\}$ and $\hat m_J$ is the no-click operator for all detection channels in $\mathcal{I}_J$, $\hat m_{\mathcal{I}_{J}}{=}\prod_{j\in \mathcal{I}_{J}}\hat m_j$.
	A full partition of the non-click operators yields the condition 
	\begin{align}\label{eq:full}
		\langle {:} \hat m_1 \dots \hat m_N {:}\rangle - \langle {:} \hat m_1 {:}\rangle \cdots \langle {:} \hat m_N {:}\rangle \stackrel{\mathrm{cl}}\geq 0.
	\end{align}

\subsection{Independence of uncorrelated noise contributions}

	Furthermore, we can show that the derived covariance conditions do not depend on uncorrelated noise contributions such as detector dark counts.
	Let us assume that two detectors have linear detector responses which can be described by the functions $\hat\Gamma_{i(j)}(\hat n_{i(j)})=\eta_{i(j)}\hat n_{i(j)}+\nu_{i(j)}$, where $\eta_{i(j)}$ and $\nu_{i(j)}$ are the quantum efficiencies and the uncorrelated noise-count rates, respectively.
	In this case, the normal-ordered expectation values of the no-click operators are $\langle{:}\hat m_i \hat m_j{:}\rangle=e^{-(\nu_i+\nu_j)}\int d^2\alpha P(\alpha) e^{-\eta_i|u_i|^2|\alpha|^2}e^{-\eta_j|u_j|^2|\alpha|^2}$ and $\langle{:}\hat m_{i(j)} {:}\rangle=e^{-\nu_{i(j)}}\int d^2\alpha P(\alpha) e^{-\eta_{i(j)}|u_{i(j)}|^2|\alpha|^2}$.
	As above, we can now derive an inequality which can be written in the form
	\begin{align}
		e^{-(\nu_i+\nu_j)}\left[\int d^2\alpha P(\alpha) e^{-\eta_i|u_i|^2|\alpha|^2}e^{-\eta_j|u_j|^2|\alpha|^2}-\int d^2\alpha P(\alpha) e^{-\eta_{i}|u_{i}|^2|\alpha|^2}\int d^2\alpha P(\alpha) e^{-\eta_{j}|u_{j}|^2|\alpha|^2}\right]\stackrel{\mathrm{cl}}\geq0
	\end{align}
	This can be further simplified to 
	\begin{align}
		e^{-(\nu_i+\nu_j)}\langle{:}\mathrm{Cov}(\hat m_i, \hat m_j){:}\rangle_{\mathrm{FUN}}\stackrel{\mathrm{cl}}\geq0,
	\end{align}
	where $\langle{:}\mathrm{Cov}(\hat m_i, \hat m_j){:}\rangle_{\mathrm{FUN}}$ corresponds to the covariance in the case when the detected light field would be free of any uncorrelated noise (FUN).
	We see that the uncorrelated noise contributions only result in an overall scaling of the inequality but do not influence the sign of the inequality.
	Therefore, the uncorrelated noise contributions do not alter the corresponding nonclassicality verification.
	This consideration can be straightforwardly generalized to the multimode conditions.
	Note that a similar independence of dark-count contributions has been reported in the context of phase-sensitive measurements with multiplexing detectors~\cite{Lipfert15}.

\section{Radiation state and detection model for clusters of single-photon emitters}
\label{sec:model}

	We derive the model for the multiplexed detection of light from a cluster of incoherently emitting single-photon emitters.
	This model is used to explain and interpret the experimental results for the different cluster sizes presented in Fig. \ref{fig:data1} of the main text.
	We consider clusters consisting of $M$ quantum-dot emitters and we assume that all quantum dots have equal properties and that each quantum dot emits a single photon.
	Thus, the light emitted by the cluster can be expressed by a tensor-product state of single-photon states, i.e., $|\chi\rangle=\prod_{i=1}^M\otimes|1\rangle^i$ where the upper index indicates the different single-photon modes.
	Note that the no-click operators of the different detection channels are labeled with a lower index.
	In the experiment each quantum dot is excited with a certain finite probability, so that unexcited quantum dots do not contribute to the single photon emission.
	We note that a quantum dot which does not emit a photon is equivalent to one emitting a photon but the photon gets lost, i.e., it is not recorded.
	Therefore, we consider that each quantum dot emits a single photon but we assign a quantum efficiency $\eta_{\mathrm{ex}}$ in the detection of the light which accounts for the finite probability of exciting each quantum dot.
	Furthermore, not all light emitted by the clusters is collected which can be modeled by the collection efficiency $\eta_{\mathrm{col}}$.
	We assume that the multiplexed detection is symmetric and that the on-off detectors have a linear detector response which is characterized through the detector efficiency $\eta_{\mathrm{det}}$.
	All the different efficiencies can be summarized in the total efficiency of the experiment, $\eta=\eta_{\mathrm{ex}}\eta_{\mathrm{coll}}\eta_{\mathrm{det}}$.
	Furthermore, the detection in each channel is not mode sensitive as all impinging photons can lead to a detection click.
	Therefore, we can write the no-click operator of the $j$-th detection channel as $\hat m_j=\exp [\eta(\sum_{i=1}^M \hat n_j^i)]$.

	Let us now consider the multiplexing of the tensor product single-photon states.
	The multiplexing device acts equally on each single-photon state and we assume a symmetric splitting into the four detection channels which yields
	\begin{align}
		\prod_{i=1}^M\otimes|1\rangle^i\to
		\prod_{i=1}^M\otimes\left[\frac{1}{\sqrt{4}}\left(|1,0,0,0\rangle^i+|0,1,0,0\rangle^i+|0,0,1,0\rangle^i+|0,0,0,1\rangle^i\right)\right].
	\end{align}
	Now we calculate the expectation value of the no-click operator in the $j$-th detection channel to be $\langle \hat m_j \rangle=(1-\frac{\eta}{4})^M$.
	Similarly, the expectation value of the joint no-click event for all four detection channels is given by $\langle {:} \hat m_{1} \dots \hat m_{4} {:}\rangle = (1-\eta)^M$.
	This allows us to evaluate the expressions of the condition \eqref{eq:covar} and \eqref{eq:full} $(N=4)$ to be
	\begin{align}
		\langle {:} \hat m_{1} \hat m_{3} {:}\rangle - \langle {:} \hat m_{1} {:}\rangle \langle {:} \hat m_{3} {:}\rangle=(1-\frac{\eta}{2})^M-[(1-\frac{\eta}{4})^M]^2
		 \\
		\langle {:} \hat m_{1} \dots \hat m_{4} {:}\rangle - \prod_{i=1}^{4} \langle {:} \hat m_{i} {:}\rangle= (1-\eta)^M-[(1-\frac{\eta}{4})^M]^4, 
	\end{align}
	respectively, which are given as as Eqs. \eqref{eq:model1} and \eqref{eq:model2} in the main text.
	These results are used in Fig. \ref{fig:data1} of the main paper to explain the dependence of the nonclassicality conditions on the cluster size. 

	\begin{figure}[h]
		\centering
		\includegraphics[width=0.9\columnwidth]{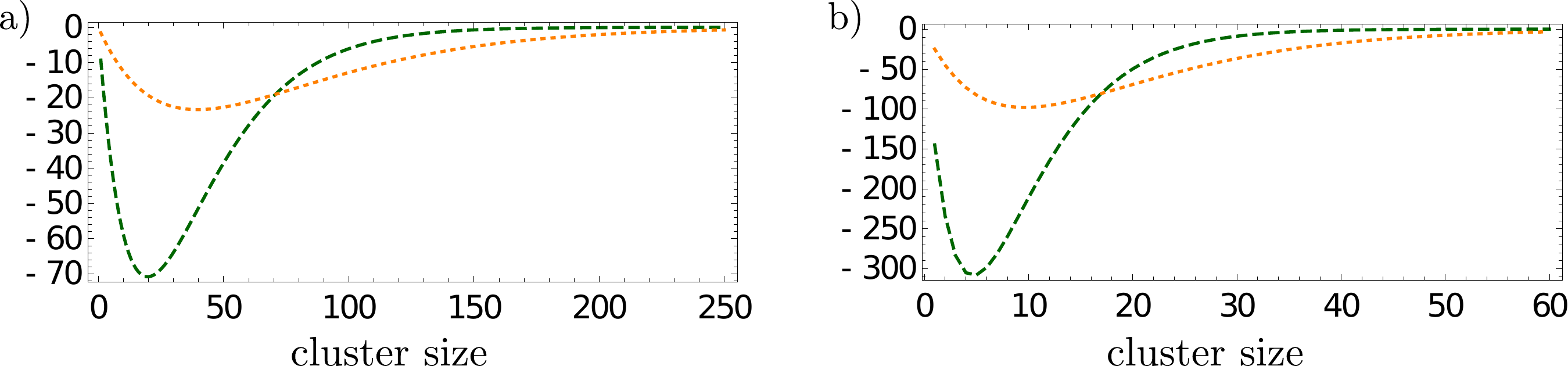}
			\caption{
				The values of the two nonclassicality conditions scaled by a factor of $10^4$ for the state and detection model, Eq. \eqref{eq:model1} (dotted) and Eq. \eqref{eq:model2} (dashed), are shown in dependence on the number of emitters for $\eta=0.05$ (a) and $\eta=0.2$ (b).
			}
		\label{fig:eta}
	\end{figure}

	Let us note that the only parameter in this model is the overall efficiency $\eta$ which accounts for all inefficiencies of the whole setup.
	For the analysis of our experimental data we estimate this efficiency to be $\eta\approx 0.009$.
	This value is in good agreement with the estimated losses and efficiencies of the experiment, and the corresponding model fits well the experimental results (cf. Fig. \ref{fig:data1}).
	The behavior of the calculated nonclassicality conditions for two other values of $\eta$ is exemplarily shown in Fig. \ref{fig:eta}.
	We observe that they all follow the same behavior, i.e., that the nonclassicality condition is at first getting more negative with increasing numbers of emitters, reaches its minimum for a certain number of emitters $M_{\mathrm{min}(\eta)}$, and eventually approaches the classical limit of zero for large numbers of emitters.
	The position of its minimal value is determined by the overall efficiency $\eta$.
	In the limit of unity efficiency $(\eta\to1)$, the minimal value is reached by one emitter ($M=1$) which is the manifestation of the single-photon character and the resulting anticorrelations in the click events (''antibunching'').
	However, we see that for lower efficiencies $\eta$  we can still certify nonclassicality even for large numbers of single-photon emitters.
	Remarkably, this is possible even if the number of detection channels ($N=4$) is orders of magnitude smaller than the numbers of emitters; cf. Fig \ref{fig:eta} and Fig. \ref{fig:data1} in the main text.
	We note that the presented model is rather simple and that it is based on several assumptions.
	This explains deviations of the model from some of the experimental data points while it describes well the overall dependence on the cluster size.

\section{Relation to other nonclassicality conditions}\label{sec:relation}
	In this section, we show some relations and differences to established methods for the verification of nonclassicality from multiplexing measurements.
	First, we will discuss the relation to conditions of the form of the Mandel $Q$ parameter.
	Second, we show that our approach delivers criteria which cannot be obtained via the matrix of moments approach.

\subsection{Relation to conditions of the form of the Mandel $Q$ parameter}
	Here, we will show how the obtained covariance condition \eqref{eq:covar} relates to established nonclassicality parameters of the form of the Mandel $Q$ parameter \cite{Mandel79}.
	Besides the Mandel parameter itself, we will also consider the recently introduced sub-binomial $Q_{\mathrm{B}}$ \cite{Sperling12a} and sub-Poisson-binomial $Q_{\mathrm{PB}}$  \cite{Lee16} parameters.
	The latter two were introduced for the verification of nonclassical light in symmetric and general (asymmetric) multiplexing schemes, respectively.
	All three parameter have in common that---whenever they attain negative values---they uncover nonclassicality of the considered quantum state.
	
\subsubsection{Sub-Poisson-Binomial parameter}
	Let us start by considering the sub-Poisson-Binomial parameter.
	The sub-Poisson-binomial parameter \cite{Lee16} may be written as
	\begin{align}\label{eq:QPB}
		Q_{\mathrm{PB}}=\frac{\sum_{i\neq j}^N \langle{:}\mathrm{Cov}(\hat m_i \hat m_j){:}\rangle}{\sum_{i}^N \langle {:}\hat m_i{:}\rangle (1-\langle{:}\hat m_i{:}\rangle)},
	\end{align}
	where $N$ is the number of detection channels and $\langle{:}\mathrm{Cov}(\hat m_i \hat m_j){:}\rangle$ is the normal-ordered expectation value of the covariance defined in Eq. \eqref{eq:covar}.
	$Q_{\mathrm{PB}}<0$ reveals nonclassicality and its sign solely depends on the numerator in Eq. \eqref{eq:QPB} as the denominator is always positive.
	Hence, $Q_{\mathrm{PB}}$ characterizes nonclassicality if the sum over all possible $\langle{:}\mathrm{Cov}(\hat m_i \hat m_j){:}\rangle$ is negative and, thus, it is based on the violation of the simple classical covariance condition \eqref{eq:covar}.
	However, it is a more complex condition as it requires evaluation of all possible correlations between the different detection channels.
	
\subsubsection{Sub-Binomial parameter}
	If we now assume that the multiplexing is performed with an equal splitting into the channels and each channel is detected with detectors which have equal properties, i.e., having the same detector response (quantum efficiency and dark-count rate), the $Q_{\mathrm{B}}$ parameter \cite{Sperling12a},
	\begin{align}\label{eq:QB}
		Q_{\mathrm{B}}=\frac{(N-1)\langle{:}\mathrm{Var}(\hat m){:}\rangle}{\langle {:}\hat m{:}\rangle (1-\langle{:}\hat m{:}\rangle)},
	\end{align}
	may be applied.
	A negative value of $Q_{\mathrm{B}}$ uncovers nonclassicality.
	Note that $Q_{\mathrm{B}}$ is the special form of $Q_{\mathrm{PB}}$ in the case of equal splitting and detection.
	In this case, all no-click operators $\hat m_i$ are equal and we can replace them by $\hat m$. 
	Then, the covariance in Eq. \eqref{eq:QPB} reduces to the variance in Eq. \eqref{eq:QB}.
	Still the negativities of $Q_{\mathrm{B}}$ arise from a negative variance, which is nothing else as the violation of the classical covariance condition \eqref{eq:covar} for the case in which all $\hat m_i$ are the same.
	The negativity of the variance is also a direct indication that the corresponding quantum state cannot be described by a classical (non-negative) $P$ function.
	Note, however, that $Q_{\mathrm{B}}$ is only applicable if the assumptions of equal splitting and detection are fulfilled.

\subsubsection{Mandel parameter}
	In \cite{Sperling12a}, it has been shown that $Q_{\mathrm{B}}$ approaches the Mandel $Q$ parameter 
	\begin{align}\label{eq:Q}
		Q=\frac{\langle{:}\mathrm{Var}(\hat n){:}\rangle}{\langle{:} \hat n{:}\rangle},
	\end{align}
	in the case of an infinite number of detection channels, i.e., $\lim_{N\to\infty} Q_{\mathrm{B}}=Q$.
	In other words, for an equal splitting into an infinite number of on-off detectors the variance of the no-click operator yields the normal-ordered expectation value of the variance of the photon-number operator, $\langle{:}\mathrm{Var}(\hat n){:}\rangle$.
	Hence, $Q$ can be seen as the limiting case of recording light with an infinite number of equal on-off detectors.
	This finding agrees with a derivation of the photon-counting formula in the 1960s \cite{Scully69}, where a bulk material described by an infinite number of single atoms (acting in the same way as on-off detectors) detects the light.
	
	We can summarize that criteria of the form of the parameters $Q$, $Q_{\mathrm{B}}$, and $Q_{\mathrm{PB}}$ are in the end based on the covariance condition in Eq. \eqref{eq:covar}.
	Importantly, the condition in Eq. \eqref{eq:covar} is the most simple form of such conditions and provides the essential building block for the other criteria.
	Moreover, the simple covariance condition using only two on-off detectors can be more sensitive than the condition provided by the $Q$ parameter, as we show in Fig. \eqref{fig:mpats},  for the example of a photon-added thermal state.

\subsection{Relation to criteria based on the matrix of moments approach}
	Here, we compare the obtained conditions with the matrix of moments approach for multiplexing detection~\cite{Sperling13}.
	For classical states, the matrix of moments $\boldsymbol M$ is positive semidefinite, with
	\begin{align}
		0\stackrel{\mathrm{cl}}\leq& \boldsymbol M=\left(\langle{:} \hat m^{l+l'}{:}\rangle\right)_{l,l'=0}^{\lfloor N/2\rfloor}.
	\end{align}
	where the floor function yields $\lfloor N/2 \rfloor=N/2$ for even $N$ and $\lfloor N/2 \rfloor=(N-1)/2$ for odd $N$. 
	A violation of this positive semidefiniteness would imply that the measured quantum state is a nonclassical one.
	As an example we can consider the $2\times2$ matrix with moments up to the $2l$th order $(l\geq1)$
	\begin{align}
		\det \begin{pmatrix}
			\langle{:} \hat m^{0}{:}\rangle & \langle{:} \hat m^{l}{:}\rangle\\
			\langle{:} \hat m^{l}{:}\rangle & \langle{:} \hat m^{2l}{:}\rangle
		\end{pmatrix}
		=\langle{:}\mathrm{Var}(\hat m^l){:}\rangle \stackrel{\mathrm{cl}}\geq 0.
	\end{align}
	This, in fact, corresponds to the correlation conditions derived here if all channels are recorded equally.
	It is, however, important to mention that, contrary to the conditions presented here, the matrix of moments approach can only be applied if equal splitting and detection are considered.
	
	We have seen that for some cases (equal splitting and detection, and $2\times2$ matrices) the matrix of moments methods yield the same conditions as our approach based on Chebyshev's integral inequality.
	This analogy, however, does not hold in general.
	In particular, our most general conditions in Eq. \eqref{eq:partitions} include arbitrary partitions which cannot be obtained with the matrix of moments approach.
	Examples are asymmetric partitions of the form 
	\begin{align}
	\langle {:} \hat m^k {:}\rangle - \langle {:} \hat m {:}\rangle \langle {:} \hat m^{k-1} {:}\rangle \stackrel{\mathrm{cl}}\geq 0\quad \text{with}\quad k>2
	\end{align}
	or multi-partitions 
	\begin{align}
	\langle {:} \hat m^k {:}\rangle - \langle {:} \hat m {:}\rangle \langle {:} \hat m^{k-2} {:}\rangle\langle {:} \hat m {:}\rangle  \stackrel{\mathrm{cl}}\geq 0\quad \text{with}\quad k>2.
	\end{align}
	Hence, the approach presented here provides new conditions which are not covered by the matrix of moments approach.
	Therefore, these more general conditions might be able to certify nonclassicality in cases where other methods fail to do so.
\end{widetext}


\end{document}